\documentstyle[12pt]{article}
\begin{document}

\vspace*{5cm}

\begin{center}
{{GENERAL GEOMETRY AND GEOMETRY OF ELECTROMAGNETISM}}
\end{center}
\vspace*{.5cm}

\begin{center}
{\bf{Shervgi S. Shahverdiyev$^*$}}
\end{center}
\begin{center}

{\small   Institute of Physics, Azerbaijan National Academy of Sciences Baku, Azerbaijan}
 \end{center}

\vspace{1cm}

\begin{center}
{\bf{Abstract}}
\end{center}

\begin{quote}
\noindent
{\small It is shown that Electromagnetism creates geometry different from
Riemannian geometry. General geometry including  Riemannian geometry
as a special case is constructed.  It is proven that the most simplest special case of General Geometry is geometry underlying Electromagnetism.
Action for electromagnetic field and Maxwell equations are derived from
curvature function of  geometry underlying Electromagnetism. And it is shown that equation of motion for a particle interacting with electromagnetic field coincides exactly with equation for geodesics of geometry underlying Electromagnetism.}
\end{quote}

\vspace{2cm}

\noindent

\hspace{.1cm}$^*$e-mail:shervgis@yahoo.com

http://www.geocities.com/shervgis

\vspace{.5cm}

{\tiny{In this version more information to all sections is added. All the main results and formulas reman the same.  Discussion of problem of geometry and matter is added to the last section. References and affiliation are added upon request of two readers, although we do not consider posting on the Internet as a publication, but a short note of new results in Physics.}}
\newpage

\section{Introduction}
After it was realized that the underlying geometry for gravitation is Riemannian geometry the action functional
for gravitational field has been derived from curvature characteristics of Riemannian geometry, namely the Lagrangian for gravitational field is scalar curvature of Riemannian geometry. And equation for geodesics coincides with the equation of motion for a particle interacting with gravitational field \cite{e},\cite{h}.

After this discovery, many physicists and mathematicians tried to find an underling geometry for electromagnetism.
The requirements for this geometry are the following; equation of motion for a particle interacting with electromagnetic field must coincide with the equation for geodesics and Lagrangian  for electromagnetic field must be related to curvature characteristics of an underlying geometry as in the case of Riemannian geometry and gravitation.
All attempts to geometrize electromagnetism has been done in the framework of Riemannian geometry in a variety of different approaches \cite{w}-\cite{shro}.
Unfortunately, these attempts failed to satisfy the above requirements  completely \cite{p}, \cite{sh} and the problem of geometrization of electromagnetism remained open.

In the present paper we show that electromagnetism can not be geometrized in the framework of
Riemannian geometry. Therefore, for geometrization of electromagnetism we need different type of geometry.
 We construct a new geometry for this aim  and call it General Geometry.  This geometry includes already known Riemannian geometry as a special case.  We introduce notion of curvature function which serves as a source for defining curvature characteristics of  geometry.
And prove that the most simplest particular case of  General Geometry is  geometry underlying Electromagnetism.

We show that equation for geodesics in  geometry underlying electromagnetism coincides with the equation of motion for a charged particle interacting with electromagnetic field and construct Maxwell equations and action functional
for electromagnetic field from curvature function.

In the next  section we show that
electromagnetism creates geometry different from  Riemannian geometry
and therefore cannot be geometrized in the framework of  Riemannian
geometry.

We develop General Geometry including Riemannian geometry as a special case in section 3.

In section 4, we derive action functional for electromagnetic field and Maxwell equations
from curvature function of the underlying geometry.

Section 5 is devoted to
discussion of correspondence between physical properties of fields and
mathematical properties of  corresponding geometries. Here, we also discuss the problem of geometry and matter.

\section{Interacting Classical Particle}

The action for a free particle is
$$
S=\frac{1}{2}m \displaystyle\int  du \eta_{\mu\nu}x^\mu_u x^\nu_u,
$$
where $x_u=dx/du, \eta_{\mu\nu}=diag(1 -1 -1 -1)$, and $m$ is mass parameter.
Consider a particle which interacts with gravitational field
$h_{\mu\nu}$ with coupling constant $\lambda$:

\begin{equation}\label{1}
S=\frac{1}{2}m \displaystyle\int  du \eta_{\mu\nu}x^\mu_u x^\nu_u
+\frac{1}{2}\lambda\int du h_{\mu\nu}x^\mu_u x^\nu_u.
\end{equation}
We can represent (\ref{1}) as
\begin{equation}\label{1a}
S= \frac{1}{2}\int du g_{\mu\nu}x^\mu_u x^\nu_u, \quad
\end{equation}
\begin{equation}\label{1b}
g_{\mu\nu}=m\eta_{\mu\nu}+\lambda h_{\mu\nu}.
\end{equation}
This has been interpreted as a free particle moving in curved spacetime
with metric $g_{\mu\nu}$ in General Relativity.

We go one step further and consider a charged particle, with charge $q$ which
interacts with
electromagnetic field $A_\mu$:
\begin{equation}\label{2}
S=\frac{1}{2}m\displaystyle\int du \eta_{\mu\nu}x^\mu_u x^\nu_u
+ \frac{q}{c}\int du A_\mu x^\mu_u,
\end{equation}
and represent (\ref{2}) as
\begin{equation}\label{2a}
S=\frac{1}{2}\int du ^A g_{\mu\nu}x^\mu_u x^\nu_u, \quad
\end{equation}
\begin{equation}\label{2b}
^Ag_{\mu\nu}=m\eta_{\mu\nu}+
\frac{q}{c}(A_\mu f_\nu +A_\nu f_\mu ),
\end{equation}
where $f_\nu x^\nu_u=1$, c is the speed of the light.
It easy to see that $f_\nu$ are functions of $du/dx^\nu$ and therefore
spoil locality. It is clearly seen that we cannot represent
(\ref{2}) in the form of (\ref{2a}) with local functions of $x$ and
$x_u$. This indicates that electromagnetism creates geometry
different from  Riemannian one. Hence, it cannot be geometrized in
the framework of  Riemannian geometry. The appropriate geometry for
electromagnetism is presented in sections 3 and 4.

For the general case
$
S=\frac{1}{2}\int du {\bf{g_{\mu\nu}}}(x, x_u)x^\mu_u x^\nu_u
$
when ${\bf{g_{\mu\nu}}}$ are considered as a function of $x$ and $x_u$
we obtain equation of motion
\begin{equation}\label{2c}
\frac{d^2x^\nu}{du^2}G_{\lambda\nu}+ \Gamma_{\lambda, \mu\nu}x^\mu_u
x^\nu_u=0,
\end{equation}
where
$$
G_{\lambda\nu}({\bf{g_{\mu\nu}}})=\frac{1}{2}\frac{\partial^2{\bf{g_{\mu\sigma}}}}{\partial
x^\nu_u\partial x^\lambda_u}
x^\mu_u x^\sigma_u+\frac{\partial{\bf{g_{\mu\nu}}}}{\partial
x^\lambda_u}x^\mu_u+\frac{\partial{\bf{g_{\lambda\sigma}}}}{\partial x^\nu_u}
x^\sigma_u+{\bf{g_{\lambda\nu}}},\quad
$$
$$
2\Gamma_{\lambda,
\mu\nu}({\bf{g_{\mu\nu}}})=\frac{\partial^2{\bf{g_{\mu\nu}}}}{\partial x^\sigma\partial x^\lambda_u}x^\sigma_u+
\frac{\partial{\bf{g_{\lambda\nu}}}}{\partial
x^\mu}+\frac{\partial{\bf{g_{\lambda\mu}}}}{\partial x^\nu}-
\frac{\partial{\bf{g_{\mu\nu}}}}{\partial x^\lambda}.
$$
Function $G_{\lambda\nu}({\bf{g_{\mu\sigma}}})$ plays the role of a
tensor for raising and lowering indices  and $\Gamma_{\lambda, \mu\nu}$
connection.
\begin{equation}\label{2d}
G_{\lambda\nu}(^Ag_{\mu\sigma})=\eta_{\lambda\nu},\quad
G_{\lambda\nu}(g_{\mu\sigma})=g_{\lambda\nu}.
\end{equation}

Accordingly, although $^A g_{\mu\nu}$ is different from $\eta_{\mu\nu}$ the role of a tensor for lowering indices plays $\eta_{\mu\nu}$.

\section{General Geometry}

In this section we construct a new geometry.
This geometry includes  Riemannian geometry, geometry underlying Electromagnetism (see next section),
  geometry underlying a unified model of Electromagnetism and Gravitation \cite{sh},
and infinite number of geometries, physical interpretation of which is
 not known at the present time, as special cases. Because of this we call it General Geometry.
Besides mathematical applications, this new geometry has important physical applications.
We demonstrate it in the next section.

Let M be a manifold with coordinates
$x^\lambda, \lambda=1,..., n$. Consider a curve on this manifold
$x^\lambda(u)$.
Vector field
$$V=\xi^\lambda\frac{\partial}{\partial x^\lambda}$$
has coordinates $\xi^\lambda$.
In  Riemannian geometry it is accepted that
\begin{equation}\label{RG}
\frac{d\xi^\lambda}{du}=-\Gamma^{\prime\sigma}_{\lambda\nu}(x)x_u^\nu\xi^\lambda,
\end{equation}
where $\Gamma^{\prime\sigma}_{\lambda\nu}(x)$ are functions of $x$
only.

To construct General Geometry we assume that
\begin{equation}\label{g}
\frac{d\xi^\sigma}{du}=-\Gamma^\sigma_\lambda(x,
x_u)\xi^\lambda.
\end{equation}
$\Gamma^\sigma_\lambda(x, x_u)$ are general functions of $x$ and $x_u$.
The next step is to consider $x$ as a function of two parameters $u,
\upsilon$ and find
\begin{tabular}{l}
$\lim \Delta\xi^\sigma/\Delta u\Delta\upsilon.$\\
\vspace*{-0.8cm}
{\tiny $\Delta u\hspace*{-0.1cm}\to\hspace*{-0.1cm}0$}   \\
\vspace*{-0.19cm}
{\tiny$\Delta \upsilon\hspace*{-0.1cm}\to\hspace*{-0.1cm} 0 $} \\
\end{tabular}
In order to do that we need
\vspace*{0.5cm}
$$
\frac{d\xi^\sigma}{du}=-\Gamma^\sigma_\lambda\xi^\lambda, \quad
\frac{d\xi^\sigma}{d\upsilon}=-\tilde{\Gamma}^\sigma_\lambda\xi^\lambda,
$$
$$
\Gamma^\sigma_\lambda=\Gamma^\sigma_\lambda(x, x_u, x_\upsilon),\quad
\tilde{\Gamma}^\sigma_\lambda=\tilde{\Gamma}^\sigma_\lambda(x, x_u,
x_\upsilon).
$$
After simply calculations we arrive at
\begin{center}
\begin{tabular}{l}
$\lim\displaystyle \frac{\Delta\xi^\sigma}{\Delta u\Delta\upsilon}=R^\sigma_\lambda\xi^\lambda,$\vspace*{-0.1cm}
\\
\vspace*{-1cm}
{\tiny$ \Delta u\hspace*{-0.1cm}\to\hspace*{-0.1cm}0$} \vspace*{.15cm}
  \\
{\tiny$\Delta \upsilon\hspace*{-0.1cm}\to\hspace*{-0.1cm} 0$}
 \\
\end{tabular}
\end{center}
where
$$
R^\sigma_\lambda=\frac{d}{d\upsilon}\Gamma^\sigma_\lambda-\frac{d}{du}\tilde{\Gamma}^\sigma_\lambda
+\tilde{\Gamma}^\sigma_\rho\Gamma^\rho_\lambda-\Gamma^\sigma_\rho\tilde{\Gamma}^\rho_\lambda.
$$
We call $R^\sigma_\lambda$ curvature function.

Representing $\Gamma^\sigma_\lambda(x, x_u)$ as
$$
\Gamma^\sigma_\lambda(x,
x_u)=F^\sigma_\lambda(x)+\Gamma^\sigma_{\lambda\nu}(x)x_u^\nu+\Gamma^\sigma_{\lambda\nu\mu}(x)
x_u^\nu x_u^\mu+...
$$
and considering each order in $x_u$ or their combinations separately we define  a set of new geometries.
Only the first order in $x_u$ is already known Riemannian geometry.
Let us show how curvature function is related to curvature tensor in the case of Riemannian geometry.
 Let
$$
\Gamma^\sigma_\lambda(x, x_u,
x_\upsilon)=\Gamma^\sigma_{\lambda\nu}(x)x^\nu_u, \quad
\tilde{\Gamma}^\sigma_\lambda(x, x_u,
x_\upsilon)=\Gamma^\sigma_{\lambda\nu}(x)x^\nu_\upsilon.
$$
Curvature function for this case is
$$
R^\sigma_{\lambda}=R^\sigma_{\lambda\mu\nu}(x^\nu_u
x^\mu_\upsilon-x^\nu_\upsilon x^\mu_u),
$$
where
$$
R^\sigma_{\lambda\mu\nu}=\partial_\mu\Gamma^\sigma_{\lambda\nu}-\partial_\nu\Gamma^\sigma_{\lambda\mu}
+\Gamma^\sigma_{\rho\mu}\Gamma^\rho_{\lambda\nu}-\Gamma^\sigma_{\rho\nu}\Gamma^\rho_{\lambda\mu}
$$
 is the curvature tensor of Riemannian geometry.

\section{Geometry of Electromagnetism}

In this section we consider the most simplest case of General Geometry
$$
\Gamma^\sigma_\lambda(x, x_u, x_\upsilon)=F^\sigma_\lambda(x(u,
\upsilon)), \quad
\tilde{\Gamma}^\sigma_\lambda(x, x_u, x_\upsilon)=F^\sigma_\lambda(x(u,
\upsilon)),
$$
when $\Gamma^\sigma_\lambda(x,x_u)$ does not depend on $x_u$ and show that it is an underlying geometry for electromagnetism.
In order to prove that geometry  defined by
\begin{equation}\label{GE}
\frac{d\xi^\sigma}{du}=-F^\sigma_\lambda(x)\xi^\lambda
\end{equation}
with the length of a curve
$$
ds=\sqrt{\eta_{\mu\nu}dx^\mu dx^\nu}+\frac{q}{cm}A_\mu dx^\mu
$$
is an underlying geometry for electromagnetism we must show that equation of motion for a particle interacting with electromagnetic field coincides with equation of geodesics in this geometry, and Maxwell equations and  Lagrangian for electromagnetic field are related to its curvature characteristics.

Geometry defined by (\ref{GE}) has different properties than Riemannian geometry defined by
(\ref{RG}). We do not get into details here. We simply mention that in this geometry the
 notion of parallel transport is not defined. As we show in the sequel this makes it
 be underlying geometry for Electromagnetism.

To obtain  equations for geodesics we substitute $\xi^\lambda$  in (\ref{GE}) by
$x^\lambda_u$ and arrive at
$$
\frac{d^2x_\sigma}{du^2}=-F_{\sigma\lambda}(x)x^\lambda_u.
$$
This is exactly equation of motion for a charged particle moving in
electromagnetic field $A_\mu$, if we choose
 $$
F_{\sigma\lambda}=\frac{q}{cm}(\partial_\sigma A_\lambda-\partial_\lambda A_\sigma).
$$
So, the first requirement is satisfied with this choice of function $F_{\sigma\lambda}$.
In the coming paper \cite{sss}, we prove this relation between $F_{\sigma\lambda}$ and $A_\mu$.

It remains to show that Maxwell equations and  Lagrangian for electromagnetic field is related to curvature characteristics of geometry (\ref{GE}).
For this end let us find curvature function for (\ref{GE})
$$
R^\sigma_{\lambda}=R^\sigma_{\mu\lambda}(x^\mu_\upsilon-x^\mu_u),
$$
where
$$
R^\sigma_{\mu\lambda}=\partial_\mu F^\sigma_\lambda.
$$
This tensor is an analog of curvature tensor of Riemannian geometry.
After summing by two of the three indices we obtain
$$
R_\lambda=R^\mu_{\mu\lambda}=\partial_\mu F^\mu_\lambda.
$$
Vector $R_\lambda$ is an analog of Ricci tensor.
Equations $R_\lambda=0$ coincide with the Maxwell equations.
In order to construct a Lagrangian we need a scalar function. In our case we have two quantities $R_\lambda $ and $A^\lambda$.
$A^\lambda$ originates from the length of a curve (metric) as
$g_{\mu\nu}$, originates from the length of a curve (metric) in Riemannian geometry.
We can construct from $R_\lambda $ and $A^\lambda$ a Lagrangian
$$
R=A^\lambda R_\lambda =\partial_\mu(A^\lambda
F^\mu_\lambda)-\frac{1}{2}F_{\mu\lambda}F^{\mu\lambda}.
$$
This coincides with the Lagrangian of electromagnetic field up to total derivative.

We see that as in the case of Riemannian geometry and gravitation
we can find equations and action functional for electromagnetic field from
geometric characteristics of geometry underlying Electromagnetism. And equation for geodesics coincides with the equation of motion for a particle interacting with electromagnetic field.

From the geometrical point of view a charged particle interacting with electromagnetic field can be considered
as a free particle in the spacetime with the length of a curve $ds=\sqrt{\eta_{\mu\nu}dx^\mu dx^\nu}+\frac{q}{cm}A_\mu dx^\mu$
and  equation for geodesic
$$
\frac{d^2x_\sigma}{du^2}=\frac{q}{cm}(\partial_\lambda A_\sigma-\partial_\sigma A_\lambda)x^\lambda_u,
$$
where $A_\mu$ is a solution to  equation $R_\lambda=0$.

\section{Discussion}

We note that the property of  gravitational field that we can choose
reference frame where gravitational field is absent corresponds to the
property of  Riemannian geometry that we can perform change of
variables  so that the right hand side of (\ref{RG})
will be equal to zero.

 For  electromagnetic field we cannot find  a
reference frame where it is absent. This property demonstrates again that electromagnetism creates geometry different from Riemannian and corresponds to
property of (\ref{GE}) that we cannot eliminate its  right hand side
 by changing coordinates. Therefore, geometrization of electromagnetism in geometries like
Riemannian geometry, where notion of parallel transport is defined must fail.

Because of the above mentioned correspondences we see that Riemannian geometry and Geometry of Electromagnetism are well suited for gravitation and electromagnetism respectively.
It is worth expecting that we may be able to construct geometries of weak and strong interactions using correspondence
between physical properties of these interactions and mathematical properties of some unknown at the present time geometries
  along the same lines as it is done for electromagnetism.
If we know underlying geometries for weak and strong interactions we can look for a geometry
including all geometries as special cases. Then from its curvature function we may construct an action
unifying all fundamental interactions along the same lines as in the case of gravitation and electromagnetism. In this way we will be able to find electroweak model without Higgs fields and unify all known interactions.

As it follows from the results of previous section geometry underlying electromagnetism is defined by
$$
\frac{d\xi^\sigma}{du}=-\frac{q}{cm}(\partial_\sigma A_\lambda-\partial_\lambda A_\sigma)\xi^\lambda.
$$
And the length of a curve is  $ds=\sqrt{\eta_{\mu\nu}dx^\mu dx^\nu}+\frac{q}{cm}A_\mu dx^\mu$.
We see that geometry  and the length of a curve depend on characteristics of interacting particles $q$, $m$ and sources for $A_\mu$ in contrast with the case of gravitation where
geometry depends on characteristics of sources for gravitational field only.
Geometry of electromagnetism gives us a new understanding of problem of geometry and matter with conclusion that geometry is determined by interaction.
If there is no interaction geometry is flat as it follows if we choose neutral particle ($q=0$). In this case in the presence of electromagnetic field
$A_\mu$, $\frac{d\xi^\sigma}{du}=0$.

We choose $\Gamma^\sigma_\lambda(x, x_u,
x_\upsilon)=F^\sigma_\lambda(x)+\Gamma^\sigma_{\lambda\nu}(x)x^\nu_u $
for geometry underlying  unified model of Electromagnetism and
Gravitation
\cite{sh}. In this case we construct action from curvature function
which is the sum of the action for Electromagnetic field in spacetime with
(\ref{1b}) and gravitational field $g_{\mu\nu}$. This model predicts that
electromagnetic field is a source for gravitational field.

In conclusion we note that Kaluza -Klein theory gives Maxwell equations in the weak fields approximation only.
For the full theory, without any approximations, it fails to reproduce Maxwell equations.
Besides, in Kaluza-Klein theory equation for geodesics does not coincide with the equation of motion for a classical particle interacting with electromagnetic field because of charge/mass problem \cite{k}.

\end{document}